\documentclass[showpacs,psfig,aps,pra,twocolumn]{revtex4}

\newcommand{\ba}{\begin{eqnarray}}
\newcommand{\ea}{\end{eqnarray}}

\usepackage{graphicx,color}

\newcommand{\ce}{$\rm C_{60}$}
\newcommand{\cm}{$\rm C_{60}$ }

\newcommand{\be}{\begin{equation}}
\newcommand{\ee}{\end{equation}}
\newcommand{\la}{\langle}

\newcommand{\ra}{\rangle}
\newcommand{\et}{{\it et al. }}

\newcommand{\bn}{\begin{enumerate}}
\newcommand{\en}{\end{enumerate}}

\begin{document}

\title{High-order harmonic generation in solid $\rm \bf C_{60}$}

\author{G. P. Zhang$^*$} \affiliation{Department of Physics, Indiana State
 University, Terre Haute, Indiana 47809, USA}

\author{Y. H. Bai}
\affiliation{Office of Information Technology, Indiana State
  University, Terre Haute, Indiana 47809, USA}

\date{\today}

\begin{abstract}
High harmonic generation (HHG) has unleashed the power of strong laser
physics in solids. Here we investigate HHG from a large system, solid
\ce, with 240 valence electrons engaging harmonic generation at each
crystal momentum, the first of this kind. We employ the density
functional theory and the time-dependent Liouville equation of the
density matrix to compute HHG signals.  We find that under a
moderately strong laser pulse, HHG signals reach 15th order,
consistent with the experimental results from \cm plasma. The helicity
dependence in solid \cm is weak, due to the high symmetry. In contrast
to the general belief, HHG is unsuitable for band structure mapping in
\ce. However, we find a window of opportunity using a long wavelength,
where harmonics are generated through multiple-photon excitation. In
particular, the 5th order harmonic energies closely follow the
transition energy dispersion between the valence and conduction
bands. This finding is expected to motivate future experimental
investigations.
\end{abstract}


\maketitle

The discovery of Buckminsterfullerene \cite{kroto1985} represents an
era of nanoscience and nanotechonology revolution. Solid \cm is a van
der Waals molecular crystal \cite{fischer1991}. Once it is doped with
potassium atoms, $\rm K_3C_{60}$ is a superconductor with transition
temperature at 18 K \cite{hebard1991}, which can be modulated
optically \cite{mitrano2016}. A single \cm molecule has the highest
possible point group symmetry of $I_h$, among the most symmetric
molecules ever discovered. Sixty carbon atoms form 20 hexagons and 12
pentagons, with 60 double bonds of 1.46 $\rm \AA$ between hexagons and
30 single bonds of 1.40 $\rm \AA$ between the pentagons and
hexagons. These bonds support a carbon cage of diameter 7.1 $\rm
\AA$. A mixture of $sp^2$-$sp^3$ bond is distinctively different from
those found in graphite and graphene. The $\pi$ electron cloud
surrounds the carbon cage and demonstrates a high level of charge
delocalization and conjugation. Beyond those normal molecular
orbitals, superatomic molecular orbitals are also found
\cite{feng2008,zhao2009,dutton2011,ijmpb15,pra16}.  Charge
delocalization renders fullerite a strong and fast nonlinear optical
response \cite{kafafi1992,dexheimer1993}, with the third order
susceptibility of $\chi^{(3)}$ close to $10^{-12}$ esu
\cite{flom1992}. One naturally expects a strong high harmonic
generation (HHG) from \ce. Indeed the fifth order harmonic is already
observed when one uses a strong laser \cite{lindle1993}. In 2005, we
predicted a strong HHG from a \cm molecule
\cite{prl05,pra06,josa07}. Experimentally, Ganeev \et
\cite{ganeev2009a,ganeev2009b} reported a stronger harmonic signal
from \cm plasma than from graphite, which is still stronger than that
in graphene \cite{yoshikawa}. However, up to now, there has been no
theoretical investigation in solid \ce, though investigations in other
solids are intensified recently
\cite{nc18,higuchi,Tancogne-Dejean2017c,ghimire2011,prb19,jia2019,avetissian2018,avetissian2018b}.

\begin{figure}
\includegraphics[angle=0,width=0.9\columnwidth]{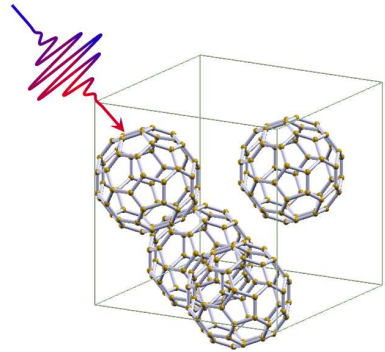}
\caption{ High harmonic generation in solid \ce. After a  fs laser pulse
  impinges on to \ce, high order harmonics are generated. Solid \cm
  crystallizes to an fcc structure at room temperature.}
\label{fig0}
\end{figure}

In this Rapid Communication, we carry out a systematic investigation
of HHG in solid \ce.  Solid \cm represents the biggest system for HHG,
with 60 carbon atoms in the unit cell (see Fig. \ref{fig0}). To handle
such a big system, we first use the density functional theory to
optimize the \cm structure. Since each unit cell has 240 valence
electrons, much larger than any prior solids \cite{ghimire2011}, the
potential violation of the Pauli exclusion principle is severe, and
the single active electron approach is clearly unsuitable. We employ
the time dependent Liouville equation for the density matrix, which
completely respects the Pauli exclusion principle.  The density matrix
is used to compute the expectation value of the momentum operator,
which is Fourier transformed to the harmonic power spectrum. We employ
two different helicities, circularly polarized and linearly polarized
laser pulses, with the photon energy from 1.6 to 2.0 eV and pulse
duration from 48 to 60 fs. We find that solid \cm has a weak helicity
dependence due to its high symmetry.  The maximum harmonic energy
under a moderately strong laser is around 20 eV, in agreement with the
experimental results \cite{ganeev2009b}. We find that in general HHG
is unable to map the relatively simple band structure of solid
\ce. However, there is a small window of opportunity if a long
wavelength of laser is used. In this case, high order harmonics need
multiple photons to lift electrons from the valence to conduction
band. In particular, the 5th order harmonics at different crystal
momentum {\bf k} do not peak at their nominal order; instead a
substantial deviation is observed with {\bf k}, an indication that HHG
involves actual band states. When we compare the transition energy
dispersion with the harmonic peak energy, we clearly see the hallmark
of the band structure. This presents a rare opportunity for HHG. We
expect this may motivate future experimental investigation.

Harmonic generation relies on a strong laser pulse.  Figure \ref{fig0}
schematically shows an example in solid \ce. High harmonics are
generated through the intense interaction between the laser and the
system.  For last thirty years, HHG investigations have focused on
atoms and small molecules. Such a spectacular development has
overshadowed the earlier work in solids \cite{farkas1992} and
nanostructures \cite{prl05}. Farkas \et \cite{farkas1992} showed that
a picosecond laser pulse can induce HHG from a gold surface.  von der
Linde \et \cite{vonderlinde1995} detected HHG up to the 15th order in
an aluminum film and the 14th order in glass where both even and odd
harmonics exist.  Faisal and Kaminski \cite{faisal1996} carried out
systematic theoretical investigations in a model thin film and
discovered harmonic orders up to 70th.  With 5 GW/$\rm cm^2$ intensity
\cite{farkas1992}, the emitted photon energy is already comparable to
the emitted energy in ZnO with TW/$\rm cm^2$ intensity
\cite{ghimire2011}. Naturally, interests in the solid state HHG are
much broader, because it potentially offers a band structure mapping
tool \cite{vampa2015a}, but this has not been materialized.

\newcommand{\rr}{{\bf r}}

\newcommand{\ik}{n{\bf k}}
\newcommand{\jk}{j{\bf k}}

Our first theoretical investigation \cite{prl05} was based on a
tight-binding model in a \cm molecule, where all the parameters are
tuned to fit the energy band gap and 174 vibrational normal mode
frequencies. Although we were limited by the model itself, the results
were already very interesting. We found that each harmonic peak can be
assigned to a specific transition among molecular orbitals. Our
present study employs the first-principles density functional theory
as implemented in the Wien2k code \cite{wien2k}. We self-consistently
solve the Kohn-Sham equation \cite{wien2k,np09,prb09}, \be \left
[-\frac{\hbar^2\nabla^2}{2m_e}+V_{ne}+V_{ee}+V_{xc} \right
]\psi_{\ik}(\rr)=E_{\ik} \psi_{\ik} (\rr), \label{ks} \ee where $m_e$
is the electron mass, the terms on the left-hand side represent the
kinetic energy, nuclear-electron attraction, electron-electron Coulomb
repulsion and exchange correlation \cite{pbe}, respectively.  We use
the generalized gradient approximation (GGA) at the Perdew, Burke and
Ernzerhof (PBE) level
\cite{pbe}.  $\psi_{\ik}(\rr)$ is the Bloch wavefunction of band $n$
at crystal momentum ${\bf k}$, and $E_{\ik}$ is the band energy.

\begin{widetext}

\begin{table}
\caption{Theoretically optimized Wykcoff positions for solid $\rm
  C_{60}$. Numbers in
  parentheses are experimental ones \cite{dorset1994}. The
  experimental lattice constant is $a=b=c=14.26 ~\rm\AA$ at room
  temperature, where \cm is disordered and the solid \cm has 
  $\rm Fm\bar{3}$ symmetry, No. 202. In our optimization, we use the
  experimental lattice constant.  }
\begin{tabular}{c|c|c|c|c|c|c}
\hline
\hline
Atom & \multicolumn{2}{c|} {$x$} &  \multicolumn{2}{c|}{$y$}& \multicolumn{2}{c} {$z$} \\
\hline
      &~~Theory~~  &Experiment     &~~Theory~~&Experiment &~~Theory~~ &Experiment\\
C$_1$ &0.04913 &(0.052)  & 0  &(0) &0.24451 &(0.249)\\
C$_2$ &0.09995 &(0.105) &0.08228 &(0.085) &0.21267& (0.220) \\
C$_3$ &0.18222 &(0.185) &0.05085 &(0.052) &0.16153& (0.165) \\
%This is K3C60 C$_1$& 96j & 0 & 0.046 & 0.245 \\
%C$_2$& 192i  & 0.213 & 0.046 & 0.245 \\
%C$_3$& 192i & 0.184 & 0.160 & 0.051 \\
\hline\hline
\end{tabular}
\label{tab}
\end{table}
\end{widetext}

Before we compute the harmonic spectrum, we first optimize the solid
\cm structure.  Different from its molecular counterpart, solid \cm
has several phases, depending on the temperature. In the room
temperature, each \cm spins rapidly and leads to a disordered phase,
with the space group symmetry of $\rm F m\bar{3}$ (No. 202)
\cite{david1991}.  We consider the room temperature phase of solid \cm
as used in the experiment \cite{ganeev2009a}.  We take the
experimental lattice constant $a=14.24~\rm \AA$ \cite{dorset1994}, and
choose a $k$ mesh of $15\times 15\times 15$.  The Wien2k code is a
full-potential augmented plane wave program. It uses dual basis
functions, atomic wavefunctions in the sphere and planewaves in the
interstitial region.  The product of the Muffin-tin radius $R_{MT}$
and the maximum planewave cutoff $K_{max}$ is 7 (dimensionless) in our
study. The optimization proceeds in two steps. First, we reduce
$R_{MT}$ by 5\% to generate a structure file, so that during the
minimization, the Muffin-tin spheres do not overlap. Second, we
minimize the total energy with respect to the carbon Wykcoff
positions. The entire minimization needs 14 iterations of structural
changes, reducing the force on each atom to around 1 mRy/a.u. No
further optimization is performed after this step. The final optimized
positions are listed in Table \ref{tab}. If we compare them with the
experimental positions (see the numbers in parenthesis), we see that
the agreement with the experimental results is quite good. Such a good
agreement for a large unit cell (60 carbon atoms) demonstrates the
accuracy of our first-principles calculation, which is further
confirmed by the band structure calculation below.

With the optimized structure in hand, we further compute the
transition matrix elements between eigenstates across the Brillouin
zone. These transition matrix elements are used for the harmonic
generation. We numerically solve the time-dependent Liouville equation
for the density matrix $\rho$ \cite{nc18,np09,prb19}, \be i\hbar
\frac{\partial \la n {\bf k}|\rho|m {\bf k}\ra}{\partial t} = \la n
     {\bf k} | [H_0+H_I, \rho] |m{\bf k}\ra, \ee where $\la n{\bf
       k}|\rho|m{\bf k}\ra$ is the density matrix element between band
     states $n$ and $m$ at ${\bf k}$ point, $H_0$ is the field-free
     Hamiltonian, and $H_I$ is the interaction Hamiltonian between the
     laser field and \ce.  Our laser vector potential has two
     different forms: Circularly polarized light ($\sigma$) and
     linearly polarized light ($\pi$). The circularly polarized light
     is described by \be {\bf A}(t)= A_0 {\rm e}^{-t^2/\tau^2}
     (\cos(\omega t) \hat{x} \pm \sin(\omega t) \hat{y}), \ee where
     $\hat{x}$ and $\hat{y}$ are the unit vectors along the $x$ and
     $y$ axes respectively, $t$ is the time, $\tau$ is the laser pulse
     duration, $\omega$ is the carrier frequency, + and $-$ refer to
     the left ($\sigma^-$) and right ($\sigma^+$) circularly polarized
     light within the $xy$ plane, respectively.  $A_0$ is the field
     amplitude in units of $\rm V fs/\AA$.  One can convert $\rm V
     fs/\AA$ to $\rm V /\AA$ using $E_0({\rm V/\AA})=A_0({\rm V
       fs/\AA})\omega$, where $\omega$ is the laser carrier angular
     frequency in 1/fs and $E_0$ is the field amplitude in $\rm V
     /\AA$.  The linearly polarized light has the following form, \be
     {\bf A}(t)= A_0 {\rm e}^{-t^2/\tau^2} \cos(\omega t) \hat{n}, \ee
     where $\hat{n}$ is the unit vector.

 The expectation value of the momentum operator \cite{prl05,pra06} is
 computed from \be {\bf P}(t)=\sum_{\bf k} {\rm Tr}[\rho_{\bf k}(t)
   \hat{{\bf P}}_{\bf k}],\ee where the trace is over band indices and
 crystal momentum {\bf k}.  The harmonic signal is computed by Fourier
 transforming ${\bf P}(t)$ to frequency domain (see the details in
 Ref. \cite{nc18}), \be {\bf P}(\Omega)=\int_{-\infty}^{\infty} {\bf
   P}(t) {\rm e}^{i\Omega t} {\cal W}(t) dt, \ee where ${\cal W}(t)$
 is the window function. ${\cal W}(t)$ has a hyper-Gaussian shape, \be
 {\cal W}_1(t)=\exp\left [-(at)^8\times b \right ], \label{gau} \ee
 where $t$ is the time in fs.  $a$ and $b$ are chosen to ensure the
 interesting regime covered. In our case, we choose $a=0.035$/fs and
 $b=5\times 10^{-9}$.  This window function is necessary since there
 are very small oscillations at the end of the time evolution. If we
 do not suppress them, they increase the overall background of the
 harmonic signal, which smears weak harmonic signals.  Our results are
 not very sensitive to $a$ and $b$ as far as the oscillatory part is
 dampened out.  We should note that using the window function does not
 affect the magnitude of harmonic signal, which is verified in our
 actual calculation.

\begin{figure}
\includegraphics[angle=0,width=0.9\columnwidth]{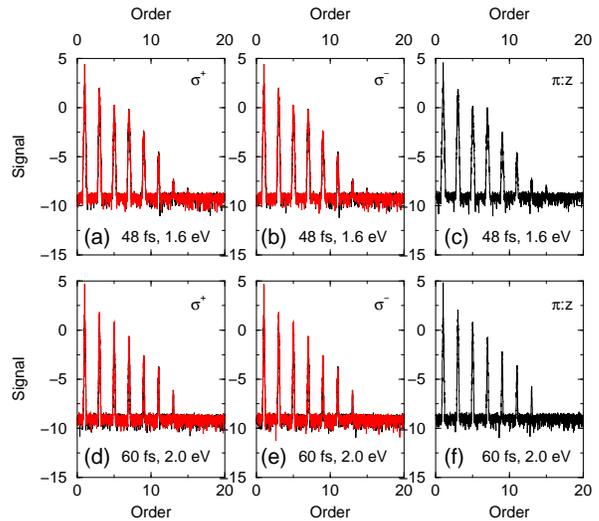}
\caption{ High harmonic generation from solid \cm under different
  laser configurations. The signal is plotted on the logarithmic
  scale.  (a) Right-circularly polarized light ($\sigma^+$) within the
  $xy$ plane.  The photon energy is $\hbar\omega=1.6$ eV, the duration
  is $\tau=48$ fs, and the laser field amplitude is $A_0=0.03~ \rm
  Vfs/\AA$. (b) Left-circularly polarized light ($\sigma^-$) within
  the $xy$ plane.  Other parameters are the same as (a).  (c) Linearly
  polarized light ($\pi$) along the $z$ direction.  Other parameters
  are the same as (a). (d) Right-circularly polarized light
  ($\sigma^+$) within the $xy$ plane.  The photon energy is
  $\hbar\omega=2.0$ eV, the duration is $\tau=60$ fs, and the laser
  field amplitude is $A_0=0.03~ \rm Vfs/\AA$. (e) Left-circularly
  polarized light ($\sigma^-$) within the $xy$ plane.  Other
  parameters are the same as (d).  (f) Linearly polarized light
  ($\pi$) along the $z$ direction.  Other parameters are the same as
  (d).  }
\label{fig1}
\end{figure}

We first choose a left-circularly polarized laser pulse ($\sigma^-$)
of $\tau=48$ fs, $\hbar\omega=1.6$ eV and $A_0=0.03 \rm
~Vfs/\AA$. Figure \ref{fig1}(a) shows the results. We see the
harmonics reach the 15th order or about 24 eV, consistent with the
experimental results \cite{ganeev2009a}. We also employ a
right-circularly polarized laser pulse ($\sigma^+$) and the results
are similar (see Fig. \ref{fig1}(b)).  This lack of helicity
dependence of HHG signal is expected from our highly symmetric system.
We also polarize our laser polarization along the $z$ axis, instead of
the $x$ and $y$ axes alone. Here the signal is only along the $z$
axis. Figure \ref{fig1}(c) shows the harmonics also reach the 15th
order.  This demonstrates that harmonics are very robust, which
explains why Ganeev \et \cite{ganeev2009a} demonstrated a much higher
(25 times) harmonic yield than carbon.

The above results only use one photon energy and one pulse
duration. We want to see whether different photon energies and
durations alter our results. We increase the pulse duration to 60 fs
and photon energy to 2 eV. Figure \ref{fig1}(d) shows the results for
$\sigma^+$. We see that for the same laser field amplitude, the
highest harmonic order is reduced to 13, so the maximum harmonic
energy is still around 26 eV.  We should add that in our calculation,
we include all the 120 valence bands (carbon has four valence
electrons: $2s^22p^2$) from -1.22 Ry and 40 conduction bands up to
0.72 Ry, so the 20-eV plasma excitation is covered. The entire
calculation has 120 electrons for each spin channel, and as pointed
out by Ganeev \et \cite{ganeev2009a}, a single active electron
approximation is highly inadequate. Our Liouville formalism handles
the Pauli exclusion principle exactly without making any approximation
\cite{jpcm16} which is not the case in the time-dependent density
functional theory.  When we use a $\sigma^-$ pulse, we find the
results remain the same (see Fig. \ref{fig1}(e)).  This is similar to
our results with a pulse of 48 fs and 1.6 eV. We also calculate the
harmonic signal along the $z$ axis (Fig. \ref{fig1}(f)) where the
maximum order is also at 13.

\begin{figure}
\includegraphics[angle=0,width=0.8\columnwidth]{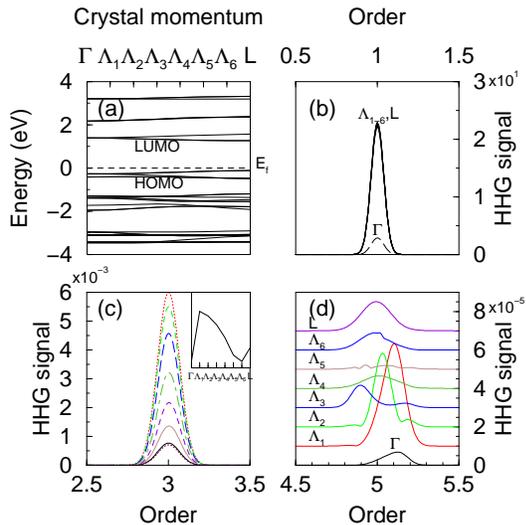}
\caption{ (a) Energy band dispersion along the $\Lambda$ line from the
  $\Gamma$ to $L$ point. (b) The first harmonic is resolved along the
  $\Lambda$ line. Here and below harmonic signals are plotted in their
  natural unit. The laser field amplitude is 0.03 $\rm Vfs/\AA$, the
  photon energy is $\hbar\omega=0.4$ eV, the pulse duration is $\tau=$
  60 fs, and the linearly polarized light along the $z$ axis is used.
  The first order harmonic has no big difference among different {\bf
    k} points.  The weakest signal is at the $\Gamma$ point.  (c) The
  third harmonic is plotted along the $\Lambda$ line. The amplitude
  change is plotted in the inset, where a clear dispersion is
  noted. (d) The fifth harmonic shows a stronger dispersion. Both the
  harmonic signal strength and the peak position disperse strongly
  with {\bf k}. All the curves, except the one at $\Gamma$, are
  vertically shifted for clarity.  }
\label{fig2}
\end{figure}

A particular question is whether one can use HHG as a band probing
tool. So far, there is no definitive answer. Vampa \et
\cite{vampa2015a} suggested to use the coherent motion of
electron-hole pairs to construct the band structure for ZnO, but it is
unclear how general this method is, in particular how one can sample
the entire Brillouin zone. We feel that here the theory can make an
important contribution.  Solid \cm has a weak energy dispersion.
Fig. \ref{fig2}(a) shows the band structure along the $\Lambda$ line
from the $\Gamma $ to $L$ point, where the Fermi level $E_f$ is
denoted by a horizontal dotted line.  The highest occupied molecular
orbital (HOMO) and the lowest unoccupied molecular orbital (LUMO) from
a single \cm molecule form a band for solid \ce. Our energy band
structure agrees with the prior calculations
\cite{xu1991,saito1991,ching1991,troullier1992,erwin1993,shirley1993}.
Theoretically, it is straightforward to compute HHG signals from one
${\bf k}$ point at a time, but it is generally unknown whether and
what difference the crystal momentum (wave vector) could make to
harmonic generation. This question is important, since if there is no
major difference in HHG across different crystal momentum points, then
trying to map the band structure by HHG is futile. Theoretical
investigation serves an important auxiliary tool to uncover extremely
challenging pockets of the Brillouin zone.

The $\Lambda$ line starts from the $\Gamma$ point with the full point
group symmetry to the $L$ point with symmetry reduced. Our first
attempt to get the band structure information is to use a pulse of
60-fs and 2.0-eV, but this does not work. Higher harmonics have a much
higher transition energy, so multiple states across several eV
contribute the signal, very hard to disentangle them. So we decide to
use a smaller photon energy of $\hbar\omega=0.4$ eV, with $\tau=60$
fs.  The first-order harmonic along the $\Lambda$ line is shown in
Fig. \ref{fig2}(b). Different from Fig. \ref{fig1} where the harmonic
signals are plotted on the logarithmic scale, the ${\bf k}$-resolved
signal is not plotted on the logarithmic scale because the relative
difference among different {\bf k} is small in solids. It is clear
that the 1st harmonic shows little difference among different {\bf k},
which explains why linear optics has no crystal momentum
resolution. The only exception is the $\Gamma$ point, where the signal
is four times weaker than the rest of {\bf k} points. Because of the
high symmetry at the $\Gamma$ point, a direct transition between the
HOMO ($H_u$) and LUMO ($T_{1u}$) is forbidden due to the parity.

The situation changes at the third harmonic. Figure \ref{fig2}(c)
shows that as we move away from the $\Gamma$ point, there is an
immediate uptake in the signal strength. The harmonic signal change is
not monotonic: It decreases first with {\bf k}, and when we are closer
to the $L$ point, there is a small increase. The inset of
Fig. \ref{fig2}(c) shows the dispersion of the harmonic signal with
{\bf k}. This result answers the above important question. If one
detects the signal along these crystal momentum directions, one indeed
can probe the band dispersion in terms of transition amplitude through
the harmonic signal change. A theoretically guided probe is amenable
to future experiments. 

To probe an energy dispersion, we move to the 5th harmonic
(Fig. \ref{fig2}(d)). While its peak amplitude drops by two orders of
magnitude, in comparison with the third order, the energy dispersion
manifests itself through the peak energy shift. The peak at $\Gamma$
does not appear at its nominal order of 5 (i. e. $5\hbar\omega=2.0$
eV), and instead now appears at 2.05 eV, close to the transition
energy of 2.44 eV between the HOMO and LUMO+1 states. Shifting away
from the nominal harmonic order is an indication that band energy
states participate in HHG through real excitation, which is already
clear in the \cm molecule \cite{prl05}. Therefore, it is instructive
to examine whether the peak energy matches the transition energy
$\Delta E$ between the valence and conduction bands.

\begin{figure}
\includegraphics[angle=0,width=0.6\columnwidth]{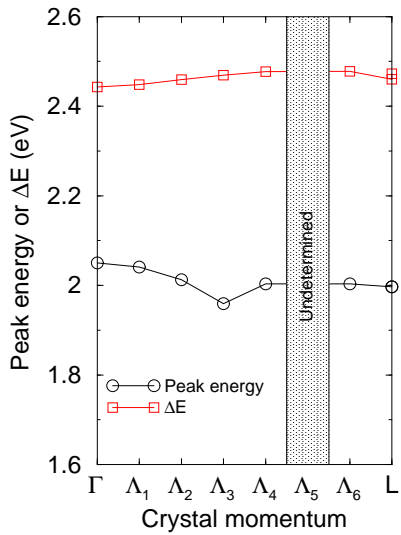}
\caption{Comparison between the harmonic peak energy of the fifth
  harmonic (the empty circles) and transition energy $\Delta E$ (the
  empty boxes) from the $H_u$ band to $T_{1g}$ band.  At the
  $\Lambda_5$, the harmonic signal is too weak, so it is not
  plotted. }
\label{fig3}
\end{figure}

Since the lowest transition is dipole-forbidden, we look at the second
lowest transition between the $H_u$ band and $T_{1g}$ band.  A true
band mapping should show that the transition energy follows the
harmonic peak energy closely.  Figure \ref{fig3} compares the fifth
harmonic peak energy and $\Delta E$.  We see that $\Delta E$ (empty
circles) is quite smooth and flat, lying between 2.4 and 2.5 eV.  The
harmonic peak energy (empty boxes), around 2.0 eV, decreases slightly
up to $\Lambda_3$, then increases and eventually saturates around 2
eV.  Since at the $\Lambda_5$ point the harmonic intensity is too
weak, we do not include it in our figure. So our conclusion is
mixed. HHG indeed has the potential to map the band
structure. However, a good match is not found in solid \ce. In
general, we expect that the flat band structure should be ideal for
HHG, but because harmonic generation involves the transition matrix
elements, just like photoemission, harmonic peak energies are
undoubtedly affected by transition matrix elements. One possible
solution is to fine tune the photon energy, but then this is not a
generic method since experimentally tuning photon energy is not easy.

In conclusion, we have investigated high harmonic generation in solid
\cm for the first time at the first-principles density functional
level. A particular challenge for solid \cm is that at each crystal
momentum point 240 valence electrons participate laser-induced
excitation, far beyond the single-active electron approach often used
in HHG calculations for atoms and small molecules.  Instead, we
employ the Liouville equation of density matrices which respects the
Pauli exclusion rigorously, in contrast to the time-dependent density
functional theory. We show that \cm generates harmonics all the way
to up to 15 with a moderately strong laser pulse, which agrees with
the prior experiment from \cm plasma \cite{ganeev2009a}. Due to the
high spherical symmetry, harmonic signals show a weak dependence on
the laser helicity. Increasing the photon energy from 1.6 to 2.0 eV,
we observe a reduction of harmonic order, which indicates there is a
maximum harmonic energy around 20 eV. This 20 eV matches the well
known plasma resonance. Solid \cm has a weak band dispersion, which is
ideal for band mapping \cite{vampa2015a}, but our results show that
there is no big difference in HHG among different crystal momenta in
general. However, if we choose a photon energy that is much smaller
than the minimum energy gap of dipole-allowed transitions between the
valence and conduction bands, higher order harmonics show a clear
dispersion.  This represents an opportunity for future experimental
investigation.

\acknowledgments

This work was solely supported by the U.S. Department of Energy under
Contract No. DE-FG02-06ER46304. Part of the work was done on Indiana
State University's high performance Quantum and Obsidian clusters.
The research used resources of the National Energy Research Scientific
Computing Center, which is supported by the Office of Science of the
U.S. Department of Energy under Contract No. DE-AC02-05CH11231.

$^*$ guo-ping.zhang@outlook.com


\begin{thebibliography}{99}
%\input{ref.tex}

\bibitem{kroto1985}H. W. Kroto, J. R. Heath, S. C. O'Brien,
R. F. Curl, and R. E. Smalley, {C$_{60}$: Buckminsterfullerene},
Nature {\bf 318}, 162 (1985).

\bibitem{fischer1991} J. E. Fischer, P. A. Heiney, A.  R. McGhie,
W. J. Romanow, A. M. Denenstein, J. P. McCauley Jr., and A. B. Smith
III, {Compressibility of Solid C$_{60}$}, Science {\bf 252}, 1288
(1991).

\bibitem{hebard1991}A. F. Hebard, M. J. Rosseinsky, R. C. Haddon,
D. W. Murphy, S. H. Glarum, T. T. M. Palstra, A. P. Ramirez, and
A. R. Kortan, {Superconductivity at 18 K in potassium-doped
C$_{60}$}, Nature {\bf 350}, 600 (1991).

\bibitem{mitrano2016} M. Mitrano, A. Cantaluppi, D. Nicoletti,
S. Kaiser, A. Perucchi, S. Lupi, P. Di Pietro, D. Pontiroli,
M. Riccò, S. R. Clark, D. Jaksch, and A. Cavalleri, {Possible
light-induced superconductivity in K$_3$C$_{60}$ at high temperature},
Nature {\bf 530}, 461 (2016).

\bibitem{feng2008}M. Feng, J. Zhao and H. Petek, {Atomlike,
hollow-core-bound-molecular orbitals of C$_{60}$}, Science {\bf
320}, 359 (2008).

\bibitem{zhao2009} J. Zhao, M. Feng, J. Yang, and H. Petek, {The
superatom states of fullerenes and their hybridization into the
nearly free electron bands of fullerites}, ACS Nano {\bf 3}, 853
(2009).

\bibitem{dutton2011} G. J. Dutton, D. B. Dougherty, W. Jin,
J. E. Reutt-Robey, and S. W. Robey, {Superatom orbitals of C$_{60}$
on Ag(111): Two-photon photoemission and scanning tunneling
spectroscopy}, \prb {\bf 84}, 195435 (2011).

\bibitem{ijmpb15}G. P. Zhang, H. P. Zhu, Y. H. Bai, J.  Bonacum,
X. S. Wu, and T. F. George, {Imaging superatomic molecular orbitals
in a C$_{60}$ molecule through four 800-nm photons}, International
Journal of Modern Physics B {\bf 29}, 1550115 (2015).

\bibitem{pra16}G. P. Zhang, A. Gardner, T. Latta, K. Drake, and
Y. H. Bai, {Superintermolecular orbitals in the C$_{60}$-pentacene
complex}, Phys. Rev. A {\bf 94}, 062501 (2016).

\bibitem{kafafi1992} Z. H. Kafafi, J. R. Lindle, R. G. S. Pong,
F. J. Bartoli, L. J. Lingg, and J. Milliken, {Off-resonant nonlinear
optical properties of C$_{60}$ studied by degenerate four-wave
mixing}, Chem. Phys. Lett. {\bf 188}, 492 (1992).

\bibitem{dexheimer1993}S. L. Dexheimer, D. M. Mittleman,
R. W. Schoenlein, W. Vareka, X.-D. Xiang, A. Zettl, and C. V. Shank,
{Ultrafast Dynamics of Photoexcited C$_{60}$}, in {\it Ultrafast
Pulse Generation and Spectroscopy}, edited by T. R. Gosnell,
A. J. Taylor, K. A. Nelson, and M. C. Downer, SPIE Proc. {\bf 1861},
328 (1993).

\bibitem{flom1992} S. R. Flom, R. G. S. Pong, F. J. Bartoli, and
Z. H. Kafafi, {Resonant nonlinear optical response of the fullerenes
C$_{60}$ and C$_{70}$}, Phys. Rev. B {\bf 46}, 15598 (1992).

\bibitem{lindle1993}J. R. Lindle, R. G. S. Pong, F. J. Bartoli, and
Z. H. Kafafi, {Nonlinear optical properties of the fullerenes
C$_{60}$ and C$_{70}$ at 1.064 $\mu$m}, Phys. Rev. B {\bf 48},
9447 (1993).

\bibitem{prl05} G. P. Zhang, {Optical high harmonic generations in
$\rm C_{60}$}, \prl {\bf 95}, 047401 (2005).

\bibitem{pra06} G. P. Zhang and T. F. George,
{Ellipticity dependence of optical harmonic generation in $\rm
C_{60}$}, \pra {\bf 74}, 023811 (2006).

\bibitem{josa07} G. P. Zhang and T. F. George, {Origin of ellipticity
anomaly in harmonic generation in $\rm C_{60}$}, J. Optical Society
of America B {\bf 24}, 1150 (2007).

\bibitem{ganeev2009a} R. Ganeev, L. Bom, J. Abdul-Hadi, M. Wong,
J. Brichta, V. Bhardwaj, and T. Ozaki, {Higher-order harmonic
generation from fullerene by means of the plasma harmonic method},
Phys. Rev. Lett. {\bf 102}, 013903 (2009).

\bibitem{ganeev2009b}R. Ganeev, L. E.  Bom, M. C. H. Wong,
J.-P. Brichta, V. Bhardwaj, P. Redkin, and T. Ozaki, {High-order
harmonic generation from $\rm C_{60}$-rich plasma}, Phys. Rev. A
{\bf 80}, 043808 (2009).

\bibitem{yoshikawa}N. Yoshikawa, T. Tamaya, and K. Tanaka,
{High-harmonic generation in graphene enhanced by elliptically
polarized light excitation},  Science {\bf 356}, 736 (2017).

\bibitem{nc18}G. P. Zhang, M. S. Si, M. Murakami, Y. H. Bai, and  T.
F. George, {Generating high-order optical and spin harmonics
from ferromagnetic monolayers}, Nat. Commun. {\bf 9}, 3031 (2018).

\bibitem{higuchi} T. Higuchi, M. I. Stockman and P. Hommelhoff,
{Stron-field perspective on high-harmonic radiation from bulk
solids}, Phys. Rev. Lett.  {\bf 113}, 213901 (2014).

\bibitem{Tancogne-Dejean2017c} N. Tancogne-Dejean, M. A. Sentef, and
A. Rubio, {Ultrafast modification of Hubbard $U$ in a strongly
correlated material: {\it ab initio} high-harmonic generation in
NiO},  Phys. Rev. Lett. {\bf 121}, 097402 (2018).

\bibitem{ghimire2011} S. Ghimire, E. Sistrunk, P. Agostini,
L. F. DiMauro, and D. A. Reis, {Observation of high-order harmonic
generation in a bulk crystal}, Nat. Phys. {\bf 7}, 138
(2011).

\bibitem{prb19} G. P. Zhang and Y. H. Bai, {Magic high-order harmonics
from a quasi-one-dimensional hexagonal solid}, Phys. Rev. B {\bf
99}, 094313 (2019).

\bibitem{jia2019}L. Jia, Z. Zhang, D. Z. Yang, M. S. Si, G. P. Zhang,
and Y. S. Liu, {High harmonic generation in magnetically-doped
topological insulators}, Phys. Rev. B {\bf 100}, 125144 (2019).

\bibitem{avetissian2018}H. K. Avetissian, A. K. Avetissian, B. R. Avchyan
and G. F. Mkrtchian, {Multiphoton excitation and high-harmonic
generation in topological insulator}, J. Phys.: Condens. Matter
{\bf 30} 185302 (2018).

\bibitem{avetissian2018b} H. K. Avetissian and G. F. Mkrtchian,
{Impact of electron-electron Coulomb interaction on the high
harmonic generation process in graphene}, Phys. Rev. B {\bf 97},
115454 (2018).

\bibitem{farkas1992} Gy. Farkas, Cs. T\'{o}th, S. D. Moustaizis,
N. A. Papadogiannis, and C. Fotakis, {Observation of
multiple-harmonic radiation induced from a gold surface by
picosecond neodymium-doped yttrium aluminum garnet laser pulses},
Phys. Rev. A {\bf 46}, R3605 (1992).

\bibitem{vonderlinde1995} D. von der Linde, T. Engers, G. Jenke,
P. Agostini, G. Grillon, E. Nibbering, A. Mysyrowicz and
A. Antonetti, {Generation of high-order harmonics from solid
surfaces by intense femtosecond laser pulses}, Phys. Rev. A {\bf    52}, R25 (1995).

\bibitem{faisal1996} F. H. M. Faisal and J. Z. Kami\'{n}ski, {Generation
and control of high harmonics by laser interaction with transmission
electrons in a thin crystal}, Phys. Rev. A {\bf 54}, R1769 (1996).

\bibitem{vampa2015a}G. Vampa, T. J. Hammond, N. Thire, B. E. Schmidt,
F. Legare, C. R. McDonald, T. Brabec, D. D. Klug, and P. B. Corkum,
{All-optical reconstruction of crystal band structure},
Phys. Rev. Lett.  {\bf 115}, 193603 (2015).

\bibitem{wien2k} P. Blaha, K. Schwarz, G. K. H. Madsen, D. Kvasnicka,
and J. Luitz, WIEN2k, An Augmented Plane Wave + Local Orbitals
Program for Calculating Crystal Properties (Karlheinz Schwarz,
Techn. Universit\"at Wien, Austria, 2001).

\bibitem{np09}G. P. Zhang, W. H\"ubner, G. Lefkidis, Y. Bai, and
T. F. George, {Paradigm of the time-resolved magneto-optical Kerr
effect for femtosecond magnetism}, {Nat. Phys.} {\bf 5}, 499
(2009).

\bibitem{prb09} G. P. Zhang, Y. H. Bai, and T. F. George, {Energy- and
crystal momentum-resolved study of laser-induced femtosecond
magnetism}, Phys. Rev. B {\bf 80}, 214415 (2009).

\bibitem{pbe}J. P. Perdew, K. Burke, and M. Ernzerhof, {Generalized
gradient approximation made simple}, Phys. Rev. Lett. {\bf 77}, 3865
(1996).

\bibitem{david1991}W.  I. F. David, R.  M. Ibberson, J.
C. Matthewman, K.  Prassides, T. John S. Dennis, J.  P. Hare, H.
W. Kroto, R.  Taylor, and D. R. M. Walton, {Crystal structure and
bonding of ordered C$_{60}$}, Nature {\bf 353}, 147(1991).

\bibitem{dorset1994}D. L. Dorset and M. P. McCourt, {Disorder and the
molecular packing of C$_{60}$ buckminsterfullerene: a direct
electron-crystallographic analysis}, Acta Crystallographica A {\bf
50}, 344 (1994).

\bibitem{jpcm16}G. P. Zhang, Y. H. Bai, and T. F. George, {Ultrafast
reduction of exchange splitting in ferromagnetic nickel}, J. Phys.:
Condens. Mat. {\bf 28}, 236004 (2016).

\bibitem{xu1991}Y. N. Xu, M. Z. Huang, and W. Y. Ching, {Optical
properties of superconducting K$_3$C$_{60}$ and insulating
K$_6$C$_{60}$},  Phys. Rev. B {\bf 44}, 13171 (1991).

\bibitem{saito1991}S. Saito and A. Oshiyama, {Cohesive mechanims and
energy bands of solid C$_{60}$}, Phys. Rev. Lett. {\bf 66}, 2637
(1991).

\bibitem{ching1991}W. Y. Ching, M. Z. Huang, Y. N. Xu, W. G. Harter,
and F. T. Chan, {First-principles calculation of optical properties
of C$_{60}$ in the fcc lattice}, Phys. Rev. Lett. {\bf 67}, 2045
(1991).

\bibitem{troullier1992}N. Troullier and J. L. Martins, {Structural and
electronic properties of C$_{60}$}, Phys. Rev. B {\bf 46}, 1754
(1992).

\bibitem{erwin1993}S. C. Erwin, {Electronic structure of the
alkali-intercalated fullerides, endohedral fullerenes, and
metal-adsorbed fullerenes}, in {\it Buckminsterfullerenes}, edited
by W.E. Billups and M.A. Ciufolini (VCH, New York), p. 217 (1993).

\bibitem{shirley1993}E. L. Shirley and S. G. Louie, {Electronic
excitations in solid C$_{60}$: Energy gap, band dispersions, and
effects of orientational disorder}, Phys. Rev. Lett. {\bf 71}, 133
(1993).

\end{thebibliography}
\end{document}